\documentclass[preprint2]{aastex}

\shorttitle{The Narrow Line Region of the Seyfert 2 galaxy Mrk 78. An infrared view}
\shortauthors{Ramos Almeida et al.}

\begin{document}

\title{The Narrow Line Region of the Seyfert 2 galaxy Mrk 78. \\
An infrared view}

\author{C. Ramos Almeida\altaffilmark{1}, A.M. P\'{e}rez Garc\'{i}a\altaffilmark{1},
J.A. Acosta-Pulido\altaffilmark{1},
J.M. Rodr\'{i}guez Espinosa\altaffilmark{1}, R. Barrena\altaffilmark{1}, and 
A. Manchado\altaffilmark{1,2}}

\altaffiltext{1}{Instituto de Astrof\'{i}sica de Canarias (IAC), 
              C/V\'{i}a L\'{a}ctea, s/n, E-38200, La Laguna, Tenerife, Spain.
cra@iac.es, apg@iac.es, jap@iac.es, jre@iac.es, rbarrena@iac.es, and
amt@iac.es}
\altaffiltext{2}{Consejo Superior de Investigaciones Cient\'{i}ficas (CSIC), Spain}

\begin{abstract}
We report near-infrared spectroscopic data for the 
Seyfert 2 galaxy Mrk~78, taken with the LIRIS near-infrared
camera/spectrometer at the William Herschel Telescope (WHT). 
The long-slit spectra clearly show extended emission. The resolution and depth
of the near-infrared spectra allows the examination of its morphology and 
ionization regions, and a direct comparison with similarly deep visible spectra. 
The emission-line ratios obtained are used to derive the extinction towards 
the nucleus.
The detection of strong features such as [Fe II], H$_{2}$, hydrogen 
recombination  lines and the coronal [Si VI]$\lambda$1.962 line is 
used to study the kinematics and excitation mechanisms occurring in
Mrk~78, revealing that despite of the strong radio-jet interaction present 
in this object, photoionization from the active nucleus dominates the narrow
line region emission, while UV fluorescence is the source of the H$_{2}$ emission.
Lines with extended emission yield velocity distributions with an amplitude
of about 600~km~s$^{-1}$, the consequence of an eastern lobe moving away 
from us plus a western lobe with the opposite contribution. 
We used the photoionization code CLOUDY to recreate a typical narrow line region
region, to derive the ionization parameter, and to compare our spectral data 
with diagnostic diagrams. 

\end{abstract}

\keywords{active galactic nuclei --- individual Mrk 78 --- near infrared}

\section{Introduction}

The availability of advanced near-infrared (NIR) instrumentation with panoramic arrays offers the 
possibility of employing diagnostic techniques that were formerly reserved to optical spectroscopy. 
Besides, long-slit NIR spectroscopy offers the possibility of characterizing 
the velocity field in AGN, using different line species from high ionization 
emission lines (e.g., [Si VI]$\lambda$ 1.962 $\mu$m), H recombination lines, 
Fe lines (mostly triggered by shocks), molecular H lines, or even stellar 
absorption features. These constitute a plethora of diagnostic tools that 
can be used to characterize many of the relevant phenomena acting in AGN, 
from the regions close to the actual nucleus, to partially ionized zones, 
to the borders of molecular clouds, etc. 
Furthermore, dust extinction is reduced in the IR with respect to the optical 
range, thus revealing lines whose optical counterparts could be obscured.

The velocity dispersions measured in the narrow emission line region (NLR) of 
Seyfert galaxies are believed to be governed by the galaxy's 
gravitational field \citep{Whittle92,Nelson96}. However, in a minority of Seyferts, with relatively luminous
linear radio sources, there is evidence for significant additional (jet related)
acceleration \citep{Ferruit02,Veilleux02}. 
The interplay between
radio-emitting flows and the line-emitting gas is important in the 
study of active galaxies, 
especially those with observable radio jets  \citep{Cecil00,Cecil02,Ferruit99}. 
This interplay can occur in a wide range of circumstances, for instance, radio 
jets can be disrupted or 
deflected by the material in the nuclear and interstellar media. In turn, 
the jets can ablate,
shock accelerate, or destroy line-emitting clouds, or even trigger bursts of 
star formation.
Mrk~78 is a classical example of an AGN with a jet--gas interaction,
showing large bipolar flows \citep{Whittle88, pedlar}, with about half of 
the NLR flux originating in two components, having velocities of 680~km~s$^{-1}$ and 
$-480$~km~s$^{-1}$, respectively.

In this paper we present NIR long-slit data of Mrk~78 with the aim of
understanding and characterizing the mechanisms responsible for its emission
spectrum. Mrk~78 is a Seyfert 2 with galactic extinction amounting to 
$E(B-V)= 0.035$ mag \citep{Schlegel98}. Its measured redshift ($z=0.0371$) 
\citep{Michel88} gives a distance of 150 Mpc and a physical scale of 
715 pc/arcsec ($H_{0}$ = 75~km~s$^{-1}$ is used throughout this paper). 
{\itshape HST} WFPC and FOC images taken in optical broad-band filters and 
[O III]$\lambda$ 5007 emission \citep{Capetti94,Whittle04} show a complex structure 
of ionized gas  aligned with the radio axis at PA = 84$^{\circ},$
plus a dust lane 
running across the galaxy center at PA = 135$^{\circ}$ (see Fig.\ref{OIII}). 
No signs of broad emission line region (BLR) emission has been detected 
from spectropolarimetry \citep{Tran95}. 
In the  soft X-ray images, taken by HRI/{\itshape ROSAT}, Mrk~78 appears extended 
and its spectrum can be fitted by a thermal plasma plus 
galactic extinction \citep{Levenson01}. 

Section 2 describes the observations, and Section 3 presents the main results, 
covering first the nuclear emission to continue with the extended 
emission spectra and the kinematics obtained from the line profiles and line 
displacements. Section 4 discusses the extinction towards Mrk~78, and makes an 
attempt at disentangling the various ionizing mechanisms at play in Mrk~78.
Finally, Section 5 summarizes the main conclusions.

\section{Observations and Data Reduction}

\begin{figure}[!h]
\includegraphics[width=6cm,angle=90]{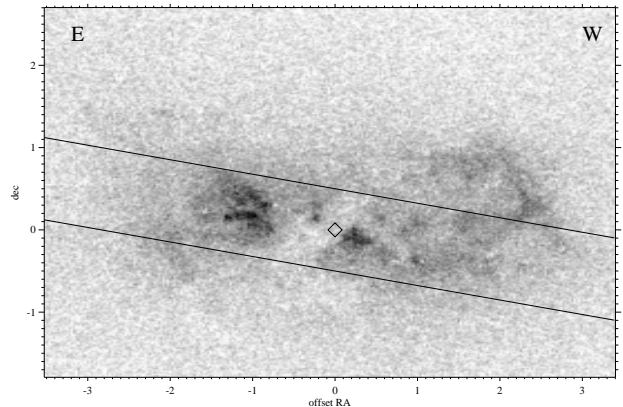}
\figcaption{\footnotesize{HST(WFPC) [OIII]$\lambda$ 5007 
image of Mrk~78, with 
the position of the LIRIS slit superimposed.
The position of the brightest IR spot is marked with a diamond.
The origin of the coordinate system is the same as reported \citet{Whittle04}}
\label{OIII}}
\end{figure}

NIR spectra in the range 0.8--2.4 $\mu$m were obtained on the night of
2005 March 22 at the 4.2 m William Herschel Telescope (WHT) using  
LIRIS, a recently 
commissioned NIR camera/spectrometer \citep{Manchado04,Acosta03}.  
LIRIS is equipped with a Rockwell Hawaii 1024$\times$1024 HgCdTe array detector. 
The spatial scale is 0.25 arcsec/pixel and the slit width used during 
the observations was
1$''$, allowing a spectral resolution of 600~km~s$^{-1}$ and 
650~km~s$^{-1}$ in the $ZJ$ and $HK$ spectra, respectively. 
The slit was oriented along PA = 80$^\circ$, centred on the galaxy 
nucleus (see Fig. \ref{OIII}). 
Weather conditions were relatively good, although with sparse cirrus. The
seeing during our observations varied bewteen 1 and $1\farcs2$, measured from
the FWHM of the comparison stars in our NIR spectra.

Observations were performed following an ABBA telescope-nodding
pattern, placing the source in two positions along the slit, separated
by 15$''$. Individual frames were taken with integration times 
of 300~s and total on-source
integration times of 60~min in each of the $ZJ$ and $HK$ spectra. 
The wavelength calibration was provided by observations of argon and xenon lamps 
available in the calibration unit at the A\&G box of the telescope.
In order to obtain the telluric correction and the flux calibration,    
the nearby HD~61232 A0 V star was observed 
with the same configuration as the galaxy.
The data were reduced following standard procedures for NIR 
spectroscopy, using
IRAF\footnote{IRAF is
distributed by the National Optical Astronomy Observatories, which
are operated by the Association of Universities for the Research in
Astronomy, Inc., under cooperative agreement with the National science
Foundation. http://iraf.noao.edu/} and the LIRIS-QL dedicated software. 
Consecutive pairs of AB two-dimensional spectra 
were subtracted to remove the sky background. The resulting frames 
were then wavelength-calibrated and flat-fielded before registering and coadding all 
frames to provide the final spectrum. 
The final wavelength-calibrated galaxy spectra were divided by a composite
spectrum to remove telluric
contamination. This composite spectrum was generated from the 
observed spectra of A0 V star HD~61232, divided by a Vega model   
convolved with the actual spectral resolution as described by \citet{vacca}.  
Differences in the strength of telluric features likely due to 
mismatch of air masses and variation of
atmospheric conditions between observations of the galaxy and the reference 
star are taken into account using Beer's law.  The IRAF task {\it telluric} 
was used in this step.
The flux calibration was carried out by normalizing with the $JHK$ magnitudes 
provided in the 2MASS survey catalogue \citep{Cutri}. The agreement in the 
continuum flux level in the overlap region for the $ZJ$ and $HK$ spectra is 
quite good in spectral shape and absolute value (about 15\%).

\section{Results}

\subsection{Nuclear Emission} 

\begin{figure*}
\centering
\includegraphics[width=9cm,angle=90]{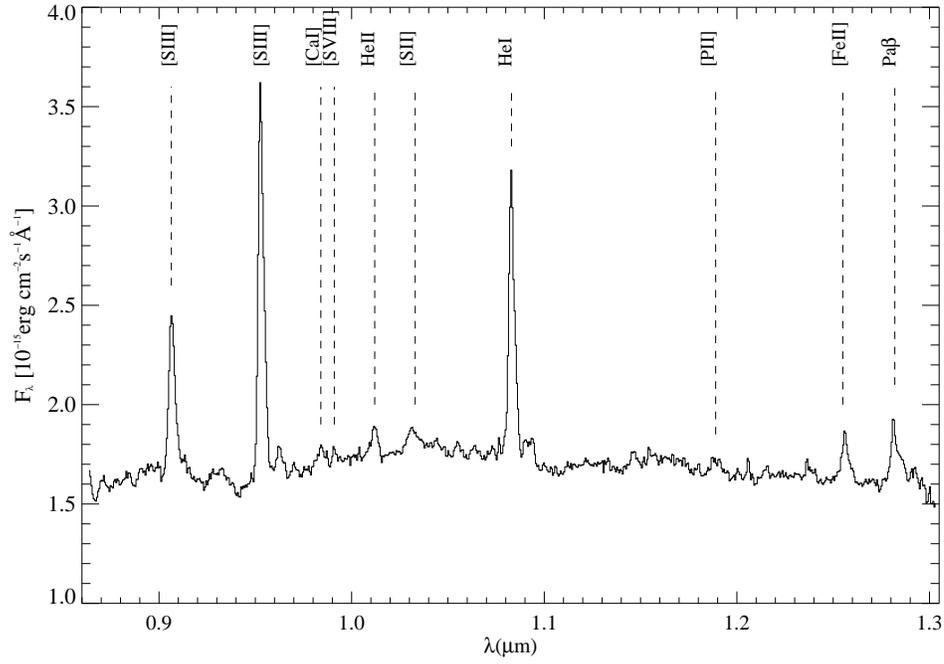}
\figcaption{\footnotesize{Flux-calibrated spectrum corresponding to
the nuclear region of Mrk 78 within an aperture of 1.25$''$, taken in the $ZJ$ 
range. Note the simultaneous detections of high and low ionization lines, 
characteristic of AGN-like spectra.} 
\label{spectrum1}}
\end{figure*}

\begin{figure*}
\centering
\includegraphics[width=9cm,angle=90]{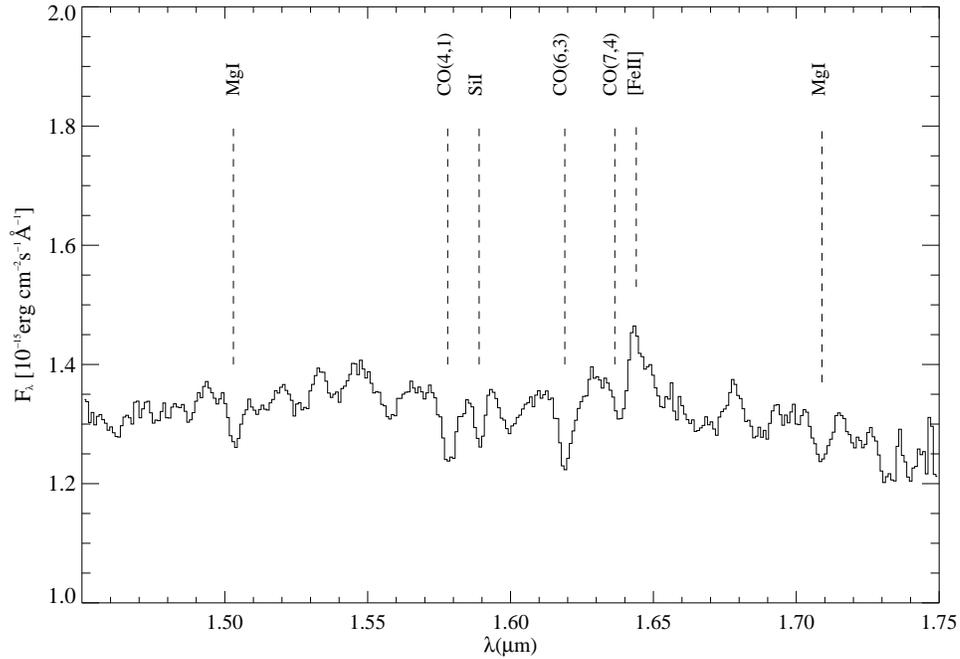}
\figcaption{\footnotesize{Same as Fig. \ref{spectrum1} but in the $H$ band. Note
the conspicuous absorption features. 
This indicates a strong stellar contribution.} \label{spectrum2}}
\end{figure*}

\begin{figure*}
\centering
\includegraphics[width=9cm,angle=90]{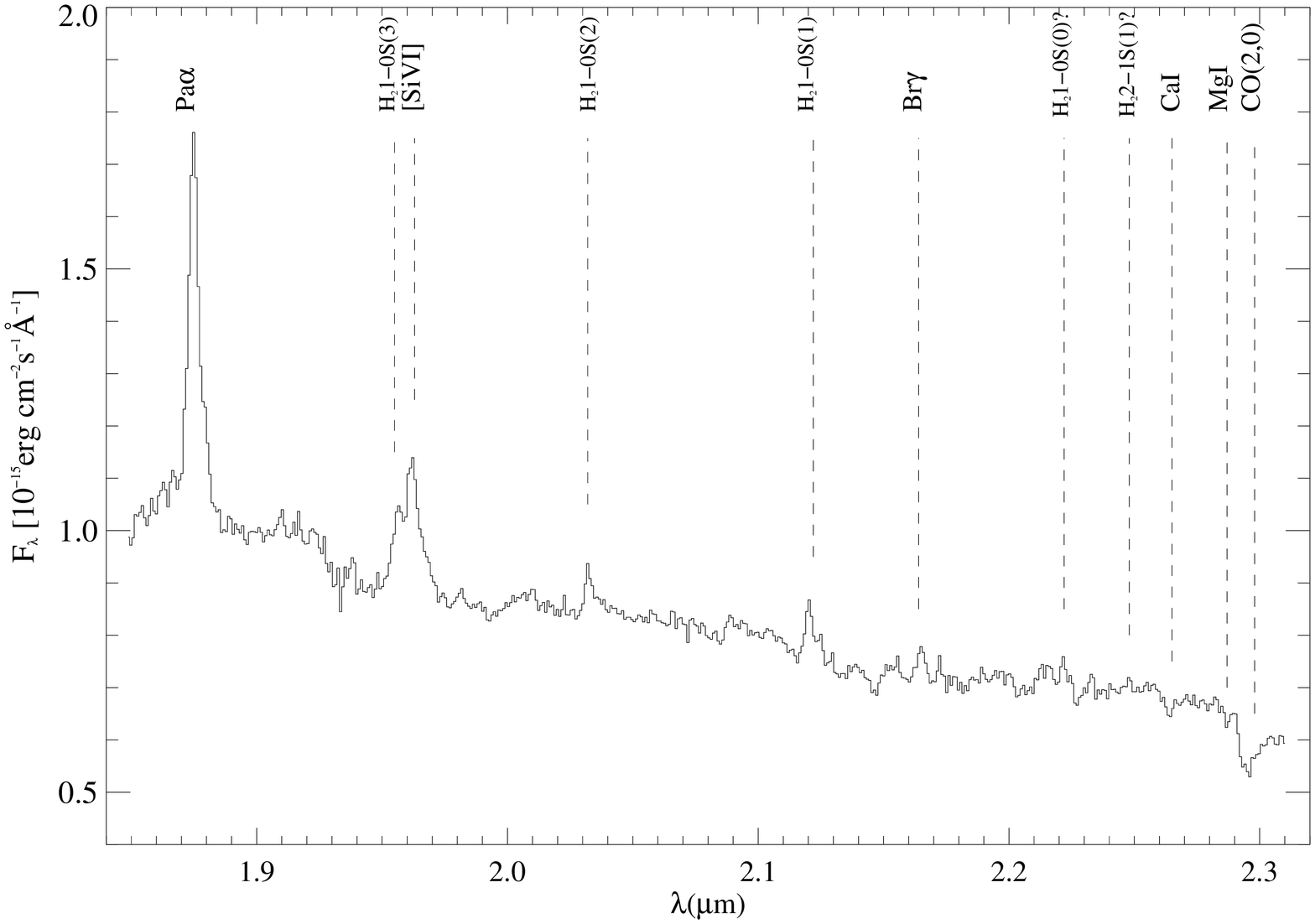}
\figcaption{\footnotesize{Same as Fig. \ref{spectrum1} but in the $K$ band. The most prominent
line in this range is Pa$\alpha$, the high ionization line [Si VI] also appears very 
intense and several H$_2$ emission lines are clearly detected.} 
\label{spectrum3}}
\end{figure*}

In order to study the nuclear emission we have extracted a spectrum 
covering 1.25~arcsec centred on the maximum of the galaxy profile.
We have identified the maximum of the continuum as the position of 
the active nucleus and registered in the 
[O III] maps (see Fig. \ref{OIII}), as  presented in \citet{Whittle04}.
These authors have done the registration of the optical (with HST 
spatial resolution) and radio images using  absolute ground--based 
astrometry \citep{Clements81}. They found a good alignment between
the peak of the optical images and the position of the radio core.
This fact gives confidence about the NIR peak should also coincide
with the position of their nucleus.   
The resulting nuclear spectra in the $ZJ$, $H$ and $K$ ranges are plotted in 
Figs \ref{spectrum1},
\ref{spectrum2}, and \ref{spectrum3}, where the wavelength has been 
translated to the observer's rest frame.
The most obvious features in the nuclear spectra are 
[S III]$\lambda\lambda$0.907, 0.953, 
He I$\lambda$1.083, [Fe II]$\lambda\lambda$1.256, 1.643, Pa$\beta$,
Pa$\alpha$, and the coronal line [Si VI]$\lambda$ 1.962 
(all wavelengths are given in $\mu$m). 
The recombination line Br$\gamma$ is also present in the $K$ band, although 
it appears very weak and is probably immersed in an absorption feature. For this
reason its measured flux is subject to a large uncertainty.
It is noteworthy to mention that in other Seyfert 2 galaxies, e.g., NGC~2273 and 
NGC~4569, this line is not found \citep{Rhee05}. 

Emission from the H$_{2}$ molecule is clearly detected, whose most 
prominent lines are the transitions H$_{2}$\,1-0\,S(2) and H$_{2}$\,1-0\,S(1). 
Both H$_2$\,2-1\,S(1) and H$_2$\,1-0\,S(0) emission lines are detected at  
 $\lesssim 2 \sigma$ level. 

Several other lower-intensity lines are detected in the $ZJ$ spectra, namely  
[Ca I]$\lambda$ 0.985, [S VIII]$\lambda$ 0.991, He II$\lambda$ 1.012,  
[S II]$\lambda$ 1.032 (a blend of 
four [S II] transitions) and [P II]$\lambda$ 1.188, which has  
particular relevance as a discriminator for different excitation mechanisms of
the gas \citep{Oliva01}.

Wherever possible all features were measured by fitting a Gaussian component
using the program {\it twofitlines} \citep{Acosta00}, developed in the IRAF
enviroment, 
and the resulting values are reported in  Table \ref{fluxes}. 
The [Si VI] and H$_2$\,1--0\,S(3) lines are strongly blended and
two Gaussians were employed to separate the emission from these lines 
(the Starlink program DIPSO was used to perform the fits). 
The resulting fluxes and EWs are also reported in Table \ref{fluxes}. 
The relative intensity between the H$_2$\,1--0\,S(3) and  H$_2$\,1--0\,S(1) lines 
is 0.83, which is in good agreement with the theoretically predicted value of 1 
for a wide range of conditions \citep{Draine90}, hence encouraging confidence 
concerning the deblending procedure. 
The [Si VI] line appears  broader ($\sim 70$~\AA ~or 1100~km~s$^{-1}$ 
after deconvolution) in comparison to other detected infrared lines
($\sim 600$~km~s$^{-1}$). 
\citet{RodriguezArdila02}  claimed a trend of increasing line widths 
with increasing ionization potential, perhaps 
indicating the existence of an emission region for the coronal lines closer 
to the nucleus than the classical NLR.

\begin{table}[!h]
\begin{center}
\caption{Observed nuclear emision lines fluxes and EWs. 
The H$_{2}$1-0S(0) and
H$_{2}$2-1S(1) fluxes reported here correspond to an upper limit of 2$\sigma$.
\label{fluxes}} 
\begin{tabular}{lcccc}
\tableline\tableline
Line Nucleus&$\lambda$ &Flux&EW\\
     &		($\mu$m) &  (10$^{-14} {\rm erg}\cdot{\rm cm}^{-2}\cdot{\rm s}^{-1})$ &(${\rm \AA}$)  \\
\tableline
$[SIII]$&	0.907& 3.54 $\pm$ 0.21& -21.5   \\
$[SIII]$&	0.953& 6.87 $\pm$ 0.21& -42.5   \\
$[CaI]$&	0.985& 0.39 $\pm$ 0.06& -3.03   \\
$[SVIII]$&	0.991& 0.17 $\pm$ 0.03&	-1.62   \\
HeII&		1.012& 0.59 $\pm$ 0.05& -3.26   \\
$[SII]$&	1.032& 0.71 $\pm$ 0.12& -4.05   \\
HeI&		1.083& 5.40 $\pm$ 0.14& -31.3   \\
$[PII]$&	1.189& 0.53 $\pm$ 0.09& -2.38   \\
$[FeII]$&	1.256& 0.96 $\pm$ 0.13& -5.93   \\
Pa$\beta$&	1.282& 1.75 $\pm$ 0.25& -10.9	\\
$[FeII]$&	1.643& 0.85 $\pm$ 0.23& -6.25	\\
Pa$\alpha$&	1.875& 4.33 $\pm$ 0.18& -40.5   \\
H$_{2}$ 1-0S(3)&1.955& 0.43 $\pm$ 0.11& -8.34   \\
$[SiVI]$&	1.962& 1.97 $\pm$ 0.17& -17.6   \\
H$_{2}$ 1-0S(2)&2.032& 0.30 $\pm$ 0.08& -3.56   \\
H$_{2}$ 1-0S(1)&2.121& 0.52 $\pm$ 0.08& -6.88   \\
Br$\gamma$&	2.165& 0.25 $\pm$ 0.02& -4.13   \\
H$_{2}$ 1-0S(0)&2.222& :0.20          &  ...    \\
H$_{2}$ 2-1S(1)&2.248& :0.20          &  ...    \\ 
\tableline
\end{tabular}
\end{center}
\end{table}

\subsubsection{Absorption-Line Features: The Stellar Population}

Several absorption features are readily seen in the $H$ band nuclear spectrum of 
Mrk~78 (see Fig. \ref{spectrum2}). In particular we observed the following ones: 
Mg I 1.50, CO 1.58, Si I 1.59, CO 1.62, CO 1.64 and Mg I 1.71 
(all wavelengths are given in $\mu$m). These features are named 
according to their main contributor, although, depending on the stellar 
spectral type, other species may become dominant. For example, for very cool 
stars, OH dominates the Si I feature \citep{Origlia93, Dallier96}. In the $K$ band, 
we have also detected  Ca I~2.26,  Mg I~2.28 ~and  
the $^{12}$CO(2,0) band head, at 2.29~\micron. Unfortunately, 
the entire band is not included in our observations 
as part of it lies outside our spectral range due to the galaxy redshift.

The ratio of the equivalent width of the 1.62 $\mu$m CO(6,3) feature to 
that of the 1.59 $\mu$m Si I one is a good temperature indicator for 
late-type stars \citep{Origlia93,Forster00,Ivanov04},
since the CO(6,3) feature grows rapidly from early K to late M stars, while the 
SiI one is only weakly dependent on stellar temperature. Moreover,
dilution and reddening effects are cancelled out, due to the closeness of the
features in wavelength.

In order to identify the spectral types that produce the absorption features 
seen in the $HK$ range of Mrk~78, we compare 
our spectra with digitally available stellar templates from \citet{Dallier96}, 
observed in the $H$ band with a medium resolution. 
The equivalent width of the intrinsic stellar features have been computed 
after convolution of the templates, in order to match the resolution of 
our galaxy spectra.   
As a result of this comparison, for Mrk~78, the equivalent-width ratio of the 
CO~1.62 feature to Si I~1.59 corresponds to spectral types in the range 
K5 III to M3 III (see Table \ref{absorption}). 
This is consistent with results obtained by \citet{Oliva95} for a range of 
galaxy types and with those of \citet{Thatte97} for NGC~1068.
An  age of several  hundred Myr is estimated for the stellar system, 
based on the fact that
intermediate to low mass giants dominate the integrated IR luminosity 
\citep{Renzini86}. 

In Seyfert galactic nuclei, these late-type stellar features are substantially 
diluted by non-stellar nuclear emission. Assuming a single stellar population, 
the  dilution fraction ($ D = 1 - {\rm EW}_{\rm obs}/{\rm EW}_{\rm int} $) can be computed 
from the ratio of  measured to  intrinsic EWs of any absorption 
feature \citep{Oliva95}.
Thus in the $H$ band, using the CO~1.62 feature, a starlight dilution  
in the range 30--37\% is obtained for the nuclear spectrum. 
For the $K$ band CO(2,0)~2.29 we use, 
as intrinsic EWs for giant stars, those computed from the 
digitally available spectra from \citet{Wallace97}, resulting in 
a  dilution factor of 43-47\%  in this band (see Table \ref{absorption}).
The stellar contribution in the $J$ band was estimated from the colors of  
K--M giants.
The remaining non-stellar spectrum resembles a power law 
($F_\nu \propto \nu^{-1.5}$).
The index of this power law is relatively close to the average value  
($F_\nu \propto \nu^{-1.35}$)
claimed for the NIR pure AGN emission \citep{neugebauer79},  
which is steeper than in the optical range. 

Another common way of determining the age of the stellar population in 
the NIR is by means of the strength of the CO absorption feature starting 
at 2.29~\micron. There are several working definitions of the
CO index. We have used that given by \citet{Ivanov00}, which is 
more appropriate to the CO band measured in Mrk~78, where the spectrum ends 
beyond 2.31~\micron. Indeed, in Mrk~78, 
because of its redshift, the CO band feature is close to the edge of 
the atmospheric window; hence, measuring the continuum level on both sides of
the feature is subject to large uncertainties.
The Ivanov CO index  is narrower than that 
of \citet{Doyon94} and has the additional advantages of being insensitive 
to extinction, and to possible uncertainties in the continuum
shape of the infrared spectra.  We  measure it as follows: 
$${\rm CO} = -2.5 \log(\langle {F_{2.295}}\rangle /\langle {F_{2.282}}\rangle ),$$

where $\langle {F_{2.282}}\rangle $ and $\langle {F_{2.295}}\rangle $ are the 
averaged flux within a bandwidth of 0.01 $\mu$m, centred on the 
blue continuum and at the band head respectively. 
The resultant CO index is 0.149 $\pm$ 0.002, which is among the lowest values 
obtained for Seyfert 2 galaxies by \citet{Ivanov00}, as expected because these 
authors use a larger spatial scale. Our value is also lower than the average 
one corresponding to the sample of pure starburst galaxies as reported by 
the same authors. 
Knowing the dilution factor from the non-stellar component in the $K$ band, 
which is approximately 45\%, we can estimate the CO index after removing the 
non-stellar contribution, being its value 0.289 $\pm$ 0.007. 
In order to estimate an age from this CO index we need 
to transform its value into an age-calibrated  one. 
First we convert our narrow spectral index into the broader, most usual one, 
using an expression from \citet{Ivanov00}:

$${\rm CO(Ivanov)}=(0.97\pm0.03)\cdot \rm CO(Doyon)+(0.002\pm0.006)$$
 
We then obtain a photometric index using a linear relationship from 
\citet{Doyon94}:

$${\rm CO_{\rm spec}}=1.46\cdot \rm CO_{\rm phot}-0.02, $$ 

which gives a CO$_{\rm phot}$ of 0.216 $\pm$ 0.023.
Looking at Fig.~1 in \citet{Origlia00}, it is ambiguous as to which isochrone is
most applicable to determine the age of the stellar population.
It is well known that the CO band strength 
is sensitive to surface gravity and depends both on metallicity and effective 
temperature. For solar metallicity, the value of the index points to $\sim$30 Myr using  
Geneva tracks \citep{Schaller92}, or to two different values, namely 
$\sim$25 and $\sim$80 Myr, if Padua tracks \citep{Bertelli94} are chosen. 
These ambiguities have led to 
the use of the CO index as an age indicator being questioned.  \citet{Origlia00} 
claim that an inadequate treatment of the AGB phase
produces  CO features that are too weak,  so that the results are fairly dependent on the 
evolutionary tracks one adopts. 
Nevertheless, if late-type giants were the dominant population in the
nucleus of Mrk~78, $\sim$100 Myr would be a lower limit to the age of the stellar
population, and  the $\sim$80 Myr point would then be the
only one possible for this case. 
Summarizing, we conclude that stellar features in the $H$ and $K$ bands are dominated 
by intermediate to low mass giants, with an estimated aged of approximately 100 Myr.

\begin{figure}[!h]
\includegraphics[width=9cm]{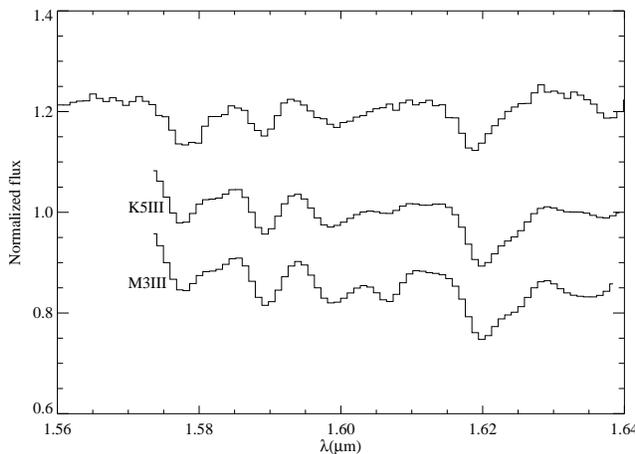}
\figcaption{\footnotesize{$H$ band spectra of Mrk~78 (top) and two giant stars (K5 III and M3 III types). 
The spectra of the stars have been convolved and rebinned in order to obtain 
the same resolution as that of the galaxy spectrum. All spectra are normalized and 
shifted vertically for clarity.}\label{bands}}
\end{figure}

\begin{table*}
\begin{center}
\caption{Absorption line feature Equivalent Widths in the H an K
bands for Mrk~78 and for three giant stars classes. (*M0III star) 
\label{absorption}}
\begin{tabular}{ccccccc}
\tableline\tableline
Line & \multicolumn{4}{c}{EW(\AA)} & Range($\mu$m)\\
& Mrk~78 & K5III & M1III & M3III\\
\tableline
SiI$\lambda$1.589& 2.3 &  2.8 &  2.9 & 2.9&1.585-1.592    \\ 
CO(6,3)$\lambda$1.619& 5.1 & 7.8 & 7.3& 8.1 & 1.614-1.627  \\
CO(2,0)$\lambda$2.290& 8.1& 14.1 & 15.2* & 15.3 & 2.291-2.303\\ 
\tableline
\end{tabular}
\end{center}
\end{table*}

\subsection{Extended Narrow Line Region}
\label{sc:extended}

Extended emission on both sides of the nucleus is observed, along the slit, 
in several spectral lines; namely, in 
[Fe II]$\lambda\lambda$ 1.256,1.643, 
[S III]$\lambda\lambda$ 0.907,0.953, He I$\lambda$ 1.083, Pa$\alpha$, 
Pa$\beta$ and [Si VI]$\lambda$ 1.962. The emission in weaker lines such as
Br$\gamma$ plus in several H${_2}$ molecular lines also 
appears extended although it becomes difficult to measure, as they are at 
about one $\sigma$ level. 
The spectral features were measured by means of Gaussian fits
using the in--house software utility {\it twofitlines} \citep{Acosta00}.
The spectra were extracted after coadding three columns
along the spatial direction, equivalent to 0$\farcs$75.
However, in a region around $1\farcs5$ east of the nucleus several lines
are split into two kinematic components (discussed in Section \ref{sc:kinem}).
In those cases the line fluxes are obtained by simple integration over the line
profile.  

\begin{figure*}
\centering
\includegraphics[width=14cm]{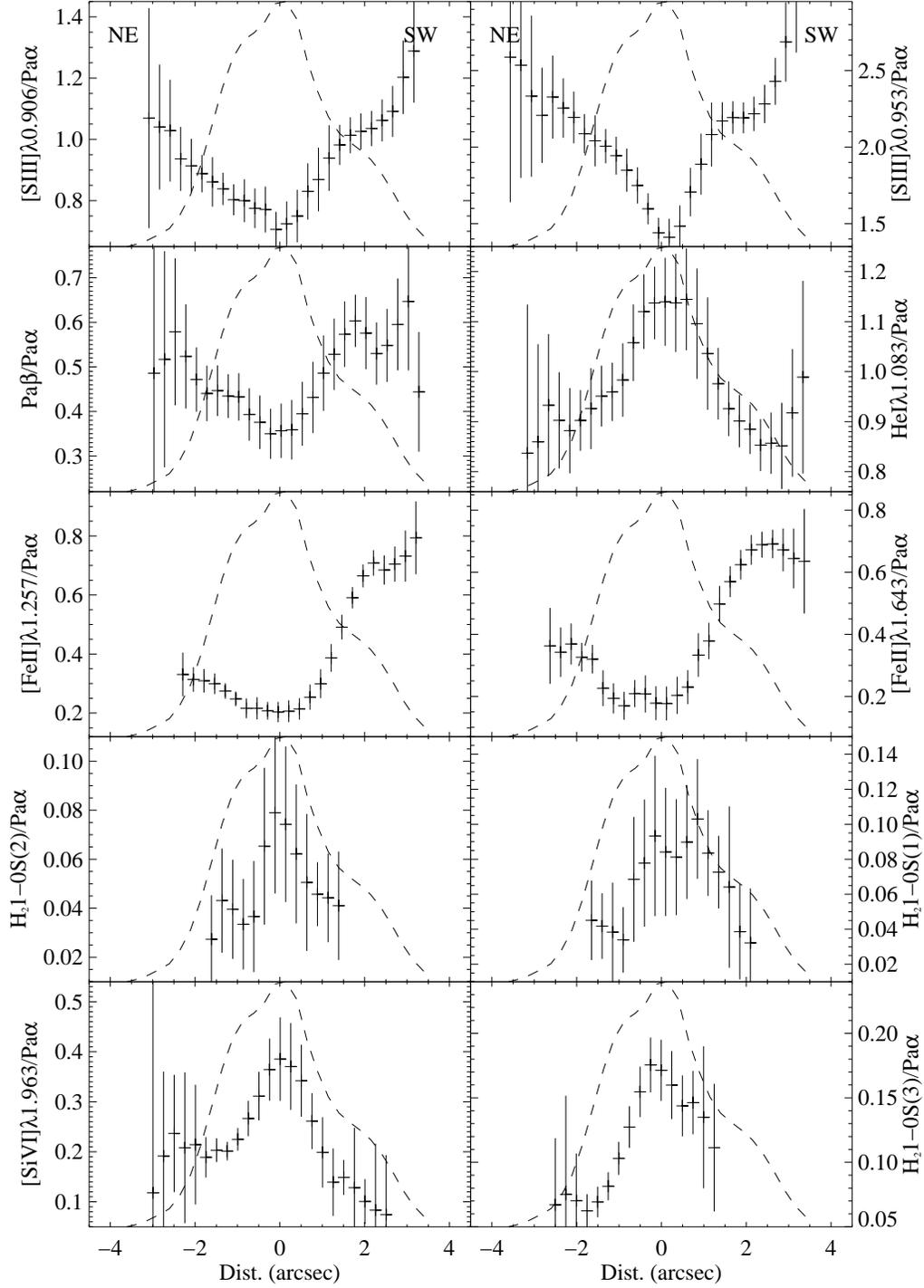}
\figcaption{\footnotesize{The most prominent emission line flux profiles vs distance 
from the nucleus, divided by the Pa$\alpha$ contribution. The dashed line represents
the flux distribution of the Pa$\alpha$ line on a logarithmic scale.}\label{fig1}}
\end{figure*}

Several line flux profiles are shown in Fig. \ref{fig1}, 
where the Pa$\alpha$ profile has been chosen as a baseline 
against which to compare the other line profiles. 
We note that the flux distribution of Pa$\alpha$ is asymmetric around the
nucleus. 
It shows first a slow decline within $1\farcs5$ east of the nucleus and then falls 
steeply beyond 2$''$. Instead, on the opposite side of the nucleus the profile falls quickly 
1$''$ west of the nucleus, further showing a shoulder at $\sim 3\arcsec.$  
The Pa$\alpha$ emission profile can be compared with the high spatial resolution
Mrk~78 [O III] and radio images from \citet{Whittle04}. 
The radio image shows two lobes on both sides of the central nucleus, which is 
actually the brightest radio source. However,
in [O III] radiation there are two 
cones of emission east and west of the center \citep{Whittle04}, which is 
consistent with the existence of a nuclear obscuring structure. 
The large size of the [O III] obscured region suggests that obscuring material 
is required on a much larger scale in addition to the few parsec size torus 
\citep{Pier92} obscuring the active nucleus \citep{Capetti94}.
A transverse [O III] profile reveals that the emission is brighter $1\farcs5$ to 
the east, decreasing towards the west
after 2$''$, just where the Pa$\alpha$ shows a shoulder.

The high-ionization line ratio [Si VI]/Pa$\alpha$ shows its maximum 
at the position of the nucleus. It also shows an enhancement $2\farcs5$ 
east of the nucleus, while it stays low ($\sim 0.1$) in
the western lobe. This is a clear indication that the ionization is 
higher in the nucleus and also in the east lobe, compared to the west side.  

In contrast, the  [S III]/Pa$\alpha$ line ratio shows a minimum 
at the nucleus and then increases to a nearly constant value in the lobes.
Both these facts are indicative of a decrease in the ionization level outside 
the nucleus, as discussed below in Section \ref{sc:cloudy}.
 
The [Fe II]/Pa$\alpha$ line ratio has an interesting behavior, showing a 
minimum at the position of the nucleus, while a high value, exceeding by a factor 
$\sim 2$ the value in the eastern lobe, appears in the 
western one. 
 
The Pa$\beta$/Pa$\alpha$ ratio is an indicative of extinction: 
there are some indications that the gas in the east and west lobes does not 
suffer so much extinction as in the nucleus, as concluded by \citet{Whittle04}.

The He I$\lambda$1.083/Pa$\alpha$ ratio  shows a shallow peak coincident with
the nucleus of the galaxy, indicating higher ionization than in the outer 
nebular regions. 

Finally, the line ratios including H$_2$ indicate that the molecular emission is
compact and concentrated towards the nucleus, although we cannot exclude the possibility that 
the molecular emission  may extend in a direction perpendicular to the 
ionization cone as mentioned by \citet{Reunanen03}.

In summary, the scenario that we may envisage through these
ratios consists of a prominent highly ionized nucleus, which is obscured by
dust; an eastern lobe where the ionization parameter is also very high
according to the [Si VI] emission (a line with a very high ionization potential);
and a western lobe where the [Fe II] emission is mostly due
to the higher radio-jet interaction in the west. 
Both lobes are less affected by extinction than the nucleus, as can be seen
from the behavior of the Pa$\beta$/Pa$\alpha$ ratio in Figure \ref{fig1}. 
This  coincides with the existence of  a dust band  crossing the galaxy center, 
as can be seen in Figure \ref{OIII}.

\subsubsection{Eastern and Western Emission}

\begin{figure}[!h]
\includegraphics[width=6cm,angle=90]{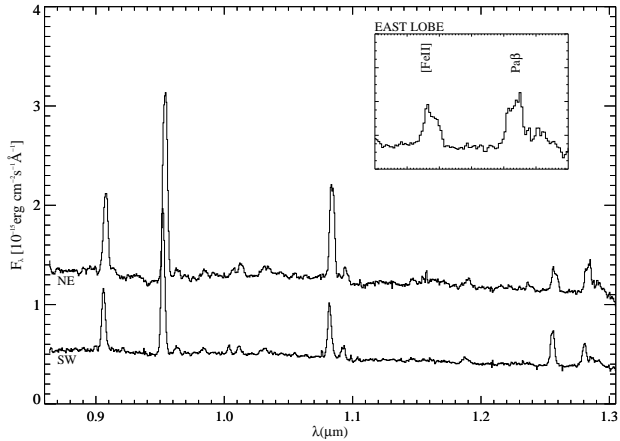}
\figcaption{\footnotesize{Flux-calibrated spectrum corresponding to $1\farcs5$ to 
the east (top spectrum) and 2$\farcs$ to the west (bottom spectrum) within an 
aperture of 1.5 arcsec, in the $ZJ$ range. The inset shows an ampliation of
[FeII]$\lambda$1.257 and Pa$\beta$ double-peaked profiles in the east lobe.}
\label{double1}}
\end{figure}

\begin{figure}[!h]
\includegraphics[width=6cm,angle=90]{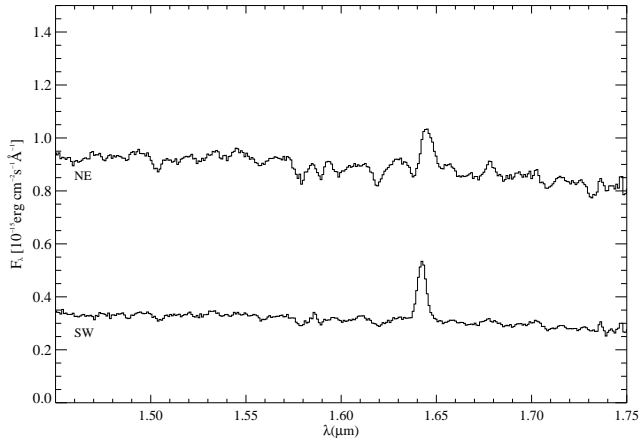}
\figcaption{\footnotesize{Same as Fig. \ref{double1} but in the $H$ band. Note the
difference in the depth of stellar features between the east and  west lobes.}
\label{double2}}
\end{figure}

\begin{figure}[!h]
\includegraphics[width=6cm,angle=90]{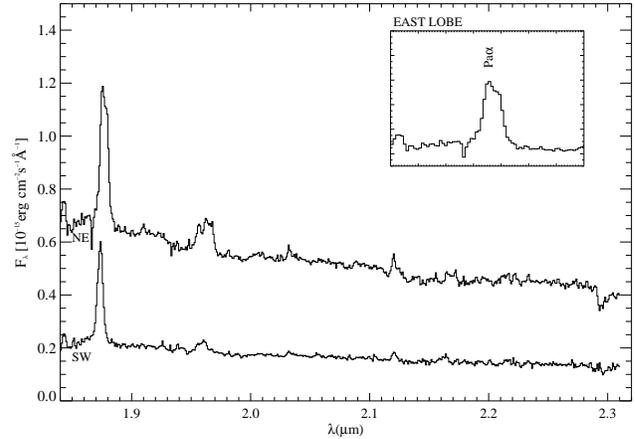}
\figcaption{\footnotesize{Same as Fig. \ref{double1} but in the $K$ band. The inset 
shows an ampliation of Pa$\alpha$ double-peaked profile in the east lobe.}
\label{double3}}
\end{figure}

By looking at the variation of the line widths along the slit, 
we noted that there are
two regions where the FWHM increases noticeably (see the discussion in 
Section~\ref{sc:kinem}). These positions coincide with the shoulders 
seen in the Pa$\alpha$ flux profile.  For this reason we decided to study
these zones separately. We extracted two spectra in the $ZJ$ and $HK$
ranges, one centred on $1\farcs5$ to the east, and the other 2$''$ to the west, 
covering 1.5 arcsec (see Figs \ref{double1}, \ref{double2}, and \ref{double3}).

In the east lobe, the most prominent lines 
(He I$\lambda$1.083, [Fe II]$\lambda$1.257, Pa$\beta$ and Pa$\alpha$) appeared  
double-peaked, as can be seen in the insets of 
Figures \ref{double1} and \ref{double3}. The two kinematic components 
correspond to a blue one, which is present overall along the slit, plus  
a weak red component, which is only detected close to  this lobe. 
In the west zone, the FWHM appears slightly larger than the mean
value, although it is not large enough to detect more than one kinematic
component. 

We report integrated fluxes for the red
and blue components of the eastern and western regions in 
Table \ref{comps}. The fluxes were measured by Gaussian fitting as described
in Section \ref{sc:extended}. In the case of the east region we employed two components, 
the velocity difference between them being $\sim 600$~km~s$^{-1}$, whereas in the west 
region a single component was used.  

The relative intensity of [Fe II] emission appears much brighter in the west; 
the values at the nucleus and at the east are similar. 
[Si VI] shows the largest value at the nucleus, and the lowest in the 
west lobe; intermediate and similar values are measured in the two components
at the east lobe. 
Differences between the blue and red spectra on the east side are
clearly seen in the  [S III] and Pa$\beta$ fluxes relative to Pa$\alpha$ 
(see Table \ref{comps}), showing a lower value for the blue component.   
	 
The depth of the stellar features appears very different in the 
east and west lobes (see Figs~\ref{double2} and \ref{double3}), being 
much larger in  the east one. 
Quantitatively, in the east the dilution factor is around
20\%, and in the west this factor is larger than 60\%, measured from the
CO(2,0) band head. Similar values are determined from  the stellar absorption 
features in the $H$ band. 
As expected for regions outside the nucleus, in the east lobe the stellar 
contribution dominates the continuum emission. In contrast, in the western lobe 
there could be an intense non-stellar continuum diluting the stellar 
counterpart, which 
could be related to the blue extended continuum claimed by \citet{Whittle04}, 
although its origin is still unclear. Otherwise, there could be an enhancement 
of old stellar population, that would contribute less to absorption features.

\begin{deluxetable}{lccccc}
\tablecaption{Nuclear, SW and NE (red and blue components) fluxes normalized to  
Pa$\alpha$ for the most prominent lines, all of them fitted by gaussians.
\label{comps}}
\tablewidth{0pt}
\tablehead{
\colhead{Line Nucleus} & \colhead{$\lambda$($\mu$m)} & \colhead{Nuclear emission} & 
\colhead{SW emission} & \colhead{Blue NE emission} & \colhead{Red NE emission}}
\startdata
$[SIII]$& 0.907 &   0.82  $\pm$   0.08    &   1.02 $\pm$   0.07    &	0.68  $\pm$   0.23   &   1.16 $\pm$   0.25	\\
$[SIII]$& 0.953 &   1.59  $\pm$   0.11	  &   2.22 $\pm$   0.13    &	1.45  $\pm$   0.11   &   2.56 $\pm$   0.16	\\
HeII&	  1.012 &   0.14  $\pm$   0.02    &   0.12 $\pm$   0.02    &	0.13  $\pm$   0.02   &   0.13 $\pm$   0.02	\\
HeI&	  1.083 &   1.25  $\pm$   0.08    &   0.89 $\pm$   0.06    &	1.16  $\pm$   0.07   &   0.96 $\pm$   0.07	\\
$[FeII]$& 1.256 &   0.22  $\pm$   0.04    &   0.61 $\pm$   0.05    &	0.29  $\pm$   0.04   &   0.22 $\pm$   0.05	\\
Pa$\beta$&1.282 &   0.40  $\pm$   0.07    &   0.45 $\pm$   0.05    &	0.32  $\pm$   0.04   &   0.51 $\pm$   0.06	\\
$[FeII]$& 1.643 &   0.20  $\pm$   0.06    &   0.63 $\pm$   0.06    &	0.29  $\pm$   0.04   &   0.15 $\pm$   0.04	\\
Pa$\alpha$&1.875&   1.00		  &   1.00     		   &    1.00  &1.00\\
$[SiVI]$& 1.962 &   0.45  $\pm$   0.06    &   0.09 $\pm$   0.04    &	0.24  $\pm$   0.02   &   0.26 $\pm$   0.03	\\
\enddata
\end{deluxetable}

\subsection{Kinematics of the Ionized Gas}
\label{sc:kinem}

A visual inspection of the two-dimensional spectrum 
reveals that the most intense emission lines describe what is most readily
explained as a rotation curve, redshifted towards the east of the galaxy 
and blueshifted towards the west. Moreover, all lines appear to be 
resolved with an average line width of 500~km~s$^{-1}$. 

\begin{figure}[!h]
\centering
\includegraphics[width=7cm]{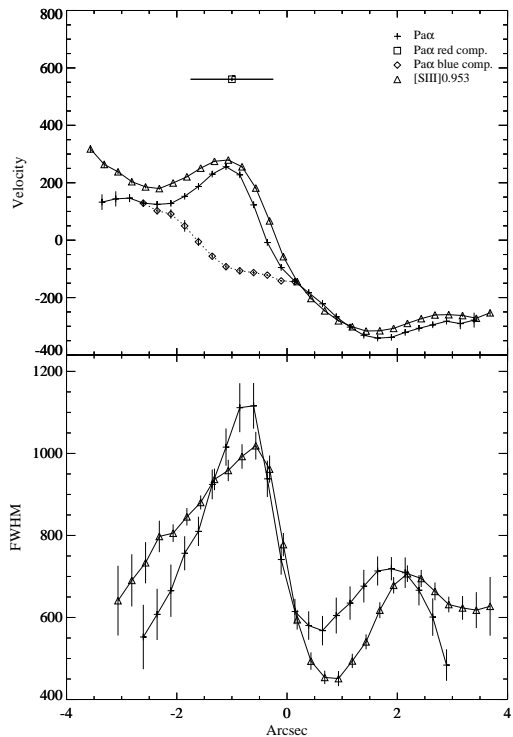}
\figcaption{\footnotesize{Variation of velocity and FWHM (both in km~s$^{-1}$) 
measured from the lines 
Pa$\alpha$ (crosses) and [S III]$\lambda 0.953$ (triangles).  
The overall measurements (fitted by a single Gaussian) are represented by 
continuous lines. 
For the Pa$\alpha$, which can be split into two components, 
the dashed line (diamonds) represents 
the blue component whereas the square indicates an average value for the 
red one.}\label{vel}}
\end{figure}

In order to analyze the velocity field we have fitted Gaussians to 
the line profiles, as described in the previous section. 
We note that around $1\farcs5$ in the
east and near 2$''$ in the west the measured line widths increase
noticeably. The velocity field appears to be
clearly perturbed in those regions. Furthermore, the presence of double-peaked 
profiles becomes
evident in some of the lines around $1\farcs5$ to the
east, e.g., He I$\lambda 1.083$, [Fe II]$\lambda 1.257$, Pa$\beta$ and Pa$\alpha$
(see Figs \ref{double1} and \ref{double3}).
We have plotted in Fig. \ref{vel} the velocity and FWHM variation along the 
slit. This has been done for an emission line where 
we clearly distinguish two components, namely Pa$\alpha$, and for another line, 
the [S III]$\lambda 0.953$, for which a single, though broader, component can be fitted.
In both cases, the overall behavior resembles that of a galaxy rotation curve with
a half amplitude of $\sim 300$~km~s$^{-1}$.  

The Pa$\alpha$ line profile, in the eastern lobe, can be split into two 
kinematic components, a blueshifted one, roughly corresponding to the
nominal rotation field, and a redshifted one, with a larger mean velocity of 
$\sim$600~km~s$^{-1}$, which may correspond to gas accelerated by interaction
with the radio jet, as proposed by \citet{Whittle04}. 

As stated above, the FWHM also increases at 2$''$ towards the west (see Fig. 
\ref{vel}), 
suggesting that the gas dynamics in this region are also affected by 
the interaction with the radio jet. However, the  spectral resolution
of our spectra does not allow us to distinguish the different kinematic components.

%_________________________________________________

\section{Discussion}

\subsection{Extinction towards the Nucleus}

According to the unification scheme for Seyfert galaxies, the nucleus of Mrk~78
should be hidden from our view. This is apparently the case, as for instance \citet{Capetti94} 
claim strong obscuration towards the nucleus of Mrk~78, based on different arguments.  
They found that the observed UV flux is at least three orders of magnitude smaller than the 
flux required to ionize the NLR. Based on this fact, applying the reddening curve from
\citet{Cardelli89} and the standard dust--to--gas ratio, they estimated a lower 
limit for the column density of $N_{\rm H} > 0.5 \times 10^{22}$~cm$^{-2}$.
They also concluded that the nuclear emission must 
be intrinsically anisotropic in order to explain the observed low 
infrared luminosity as compared to the nuclear luminosity 
required to account for the overall NLR luminosity. 
Besides, \citet{Whittle04} showed color images in which the nuclear 
extinction is clearly seen across several hundred parsecs.
Indeed, we have not detected any broad wings in the H recombination lines
present in our spectra. Previously, \citet{Goodrich94} reported the 
non-detection of a broad component for Pa$\beta$ in this galaxy, although their
spectra have much lower signal to noise ratio than ours.

Moreover, the extinction towards the nucleus can be estimated from 
our spectra using several reddening indicators, such as the 
[Fe II]$\lambda 1.257$ and [Fe II]$\lambda 1.644$ 
lines, or other line ratios between the observed Paschen and Brackett H recombination 
lines. From these line ratios 
we have determined the optical
extinction, $A_V$, using Draine's parameterization \citep{Draine89} 
$A_\lambda \propto \lambda^{-1.75}$. The results are presented in 
Table \ref{ta:extincion}.  
The  [Fe II]1.257/1.644~\micron ~ line ratio is a reliable indicator 
given that it is not dependent on the density or temperature of the line-emitting
gas. We measured a value of 1.13 (see Table \ref{ta:extincion}), 
which yields a moderate value for the 
extinction when compared to the theoretical value of 1.34 
\citep{Bautista98}. The  [Fe II]1.257/1.644 line ratio measured 
for Mrk~78 is very close to the mean value 0.98 found in a sample of 
Seyfert galaxies by \citet{RodriguezArdila04}. These authors argued 
that either the gas emission [Fe II] lines are affected by the same large 
extinction for most Seyfert galaxies, or the theoretical value 
is overestimated, given the uncertainties in the 
determination of the transition coefficients.
 
Other available reddening indicators are the 
Pa$\beta$/Pa$\alpha$ and the Pa$\alpha$/Br$\gamma$ line ratios. However, these
ratios depend on the density and temperature of the line-emitting gas. 
In particular, the Pa$\beta$/Pa$\alpha$ ratio  varies by $\sim 8$~\% when the
temperature changes from 5000 to 20000~K \citep{hummer}. 
We have adopted theoretical values corresponding to $T_e$ = 10\,000~K and 
$N_{e} = 10^4$~cm$^{-3}$. In our case, the observed Pa$\alpha$/Br$\gamma$ line ratio 
is larger than the theoretical value. This could be due to an underestimation of 
the Br$\gamma$ line flux, because of either  the weakness of the
line or  an underlying absorption feature. 

The values derived for the visual extinction agree very well with those
estimated by other authors using narrow emission lines 
\citep{Veilleux97,RodriguezArdila04}.
It is worth  mentioning that the extinction measured using these narrow lines
provides only an indication of the material located between us and the NLR. 
However, the nucleus itself and the BLR may be hidden by a larger amount 
of material. In fact, the  
non-detection of broad wings in recombination lines implies 
a lower limit of $A_v > 10$ \citep{Veilleux97}, if   
we assume the existence of a hidden BLR.

\begin{deluxetable}{lccc}
\tablecaption{Theoretical and measured emission line ratios and A$_{V}$ 
values, calculated using Draine's parametrisation.\label{ta:extincion}}
\tablewidth{0pt}
\tablehead{
\colhead{Ratio} & \colhead{Theoretical} & \colhead{Measured} & \colhead{A$_{V}$}}
\startdata
[FeII]1.26/1.64 & 1.34 & 1.13$\pm$0.34 & 1.97 \\
Pa$\beta$/Pa$\alpha$ & 0.49 & 0.40$\pm$0.07 & 1.85 \\
Pa$\alpha$/Br$\gamma$ & 12.2 & 17.3$\pm$1.6 & ...\\
\enddata
\end{deluxetable}

\subsection{The Emission-Line Spectrum}

The simultaneous observation of very high and low ionization lines imply that a 
wide variety of physical conditions must coexist in the nuclear 
region of Mrk 78. This is a characteristic feature of gas photoionized by 
a power-law radiation source extending from the UV through X-rays. In 
particular the detection of strong coronal lines such as 
[Si VI]$\lambda 1.962$ (IP = 166.7~eV) 
and [S VIII]$\lambda 0.991$ (IP = 280.9~eV) are indicative of the
presence of extreme UV and X-ray photons \citep{Prieto00}. 

\citet{Marconi94} concluded that the [Si VI]$\lambda$ 1.962 emission line 
is associated uniquely with Seyfert 1 and 2 nuclei, and can only be produced by 
photoionization by a hard nuclear continuum. Furthermore, they indicate 
that this line must be produced in a dense gas with large 
column densities. Hot stars, present in HII regions 
are not capable of exciting the [Si VI] line, neither 
increasing the ionization parameter nor the ionizing star temperature. 
This coronal line is therefore an excellent tracer of Seyfert activity 
in galaxies with visually obscured nuclei. 
Indeed, \citet{Oliva90} found strong [Si VI]  in NGC~1068  
despite the strong obscuration present in this Seyfert galaxy. 
Coronal lines are forbidden transitions of highly ionized species,
which are formed in extreme energetic environments.  
\citet{Reunanen03} observed that coronal line emission appears to be extended
along the ionization cone axis, implying an anisotropic nuclear radiation
field. Furthermore, coronal
lines are seen with similar strength and frequency in both Seyfert types
\citep{Prieto00}, thus their study provides a clean test of the Unification
Model. 

\subsubsection{The Origin of the Intense [Fe II] Emission}

It is well known that [Fe II] emission is weak in HII regions but strong
in the shock-excited filaments of supernova remnants. In AGN, strong [Fe II] 
emission is also common, although  
there is still some controversy about what is the dominant process 
responsible for this emission. Several processes
may contribute to the production of the [Fe II] lines:  
1) photoionization by soft X-ray--extreme UV radiation from the central source,
producing large partially ionized regions in NLR clouds of high optical
depth; 2) the interaction of radio jets with the
surrounding medium, which induces shocks and produces partially ionized cooling
tails; and 3) fast shocks associated with supernova remnants present in starburst
regions. 
The  [Fe II]$\lambda$1.257/Pa$\beta$ line ratio, or equivalently 
[Fe II]$\lambda$1.644/Pa$\alpha$, has proved to be very useful for distinguishing 
between a stellar and non-stellar origin for the [Fe II] emission. 
These line ratios increase from HII regions (photoionization) to supernovae 
remnants (shock excitation), passing through starburst and active galaxies 
\citep{AlonsoHerrero,RodriguezArdila04}.
Galaxies with [Fe II]$\lambda$1.644/Pa$\alpha$\footnote{We transform the  [Fe II]$\lambda$1.257/Pa$\beta$ 
line 
ratio to [Fe II]$\lambda$1.644/Pa$\alpha$ 
using the theoretical relations for both [Fe II]  and 
H Paschen lines. The following equation was used 
[Fe II]$\lambda$1.644/Pa$\alpha$ = 0.37 $\times$ [Fe II]$\lambda$1.257/Pa$\beta$.} 
lower than 0.11 are clasified as starbursts generally,
LINERS if this value is larger than 0.75, and Seyfert galaxies are contained in 
the range 0.15--0.75 \citep{Larkin98,RodriguezArdila04}. 
In the case of  Mrk~78, the [Fe II]$\lambda$1.644/Pa$\alpha$ line ratio 
is seen to increase outside the nucleus and shows larger values
in the west lobe (see Section \ref{sc:cloudy}). We found 
[Fe II]$\lambda$1.644/Pa$\alpha = 0.2$ for 
the nuclear emission and similar values in the east lobe 
that are well within the range of Seyfert galaxies. Those values can be explained 
as due to photoionization by hard UV nuclear radiation (see Fig. \ref{ioniz2}). 
Starburst activity is not sufficient to explain these values. In
the west lobe a higher value, [Fe II]$\lambda$1.644/Pa$\alpha = 0.63$, was 
measured that is closer to the LINER locus, probably related to shock 
excitation, associated with a more efficient  
interaction with the radio jet. In order to explain the 
 [Fe II]$\lambda$1.644/Pa$\alpha$ line ratio in the west lobe a composite
model including shock excitation plus photoionization by the nuclear radiation
is required.

On the other hand, we have detected [P II]$\lambda$ 1.188~\micron, 
which is an important emission line for
discriminating between the different excitation mechanisms acting in the NLR. 
In particular, this line is very useful when compared with the [Fe II]$\lambda$ 1.257
line  \citep{Oliva01}. 
Both lines are produced in partially ionized regions having  
similar critical densities and excitation temperature.
In contrast, iron is a well known refractory species and is strongly
depleted in dust grains, whereas phosphorus is a non-refractory species. 
Photoionization alone is unable to destroy the 
tough iron-based grains, while these are easily sputtered by shocks. 
The [Fe II]/[P II] ratio is high ($\ga$ 20) in fast shock-excited regions 
and low ($\la$ 2) in normal photoionized regions \citep{Oliva01}. For Mrk~78
this line ratio is 1.82 in the nuclear region, indicating that 
photoionization  by a soft X-ray continuum is the most likely mechanism 
responsible for the observed nuclear spectrum.

\subsubsubsection{The Extended [Fe II] Emission and Its Relationship to 
Radio Emission}

As pointed out by \citet{RodriguezArdila04}, the  
[Fe II]$\lambda$1.644/Pa$\alpha$ line ratio is a good 
discriminator to discern between a stellar or non-stellar origin for the 
[Fe II] emission, although it is not helpful in distinguishing between 
processes in the latter case.  
In addition, shocks may efficiently destroy dust grains and 
release iron into the
gas phase, leading to an enhancement of the [Fe II] emission.
An excellent way to differentiate between the different processes is 
the comparison of the [Fe II] emission and radio emission  morphologies, 
despite the difficulty in obtaining [Fe II] images. \citet{Blietz94} showed
that the morphology of the [Fe II] emission correlates with that of the 
radio jet in NGC~1068. The most intense [Fe II] emission 
appears slightly offset upstream of the radio bow shock (see Fig. 2 in 
\citet{Blietz94}). 
Mrk 78 may be  a similar case in which the morphology of 
the [O III]-emitting gas seems to reflect different processeses in 
the west and east lobes. The east lobe can be explained as 
a relatively homogeneous gas structure illuminated by the nuclear radiation 
source, which is penetrated by the radio emission.
In contrast, the west lobe resembles a knotty gas distribution where the 
interaction with the radio jet has dispersed cloud fragments and 
swept out gas, producing compression of the gas \citep{Whittle04}.

\subsubsection{The Origin of the H$_{2}$ Emission}

Several molecular hydrogen emission lines are present in the 
$K$ band spectrum of Mrk 78. These lines are clearly detected
in the nucleus and in the east lobe (see Figs \ref{spectrum3} and \ref{double3}), 
although the emission is mostly concentrated in the nuclear region. 
\citet{Reunanen03} 
reported that H$_2$ emission is spatially resolved preferentially in
directions perpendicular 
to the ionization cone, which is in contrast with the results reported
by \citet{Quillen99}, where H$_2$ emission coincides with [O III] and 
H$\alpha$ emission. 

There are three possible H$_{2}$ excitation mechanisms in galaxies:
i) UV pumping (fluorescence), ii) thermal excitation due to the presence of 
shocks, and iii) hard X-ray photons, which can penetrate deep into molecular
clouds.  These three mechanisms produce different spectral features, and  
the relative emission line intensities can be used to identify the dominant 
mechanism.  In particular the  H$_2$ 2--1~S(1)/1--0~S(1) line ratio is 
lower for thermal excitation (0.1--0.2) than for UV fluorescence ($\sim$0.55), 
as proposed by \citet{mouri}. The value obtained for the nuclear region of
Mrk~78 is $\la 0.38,$ which is larger than the predictions for pure thermal 
excitation, but smaller than those for non-thermal UV excitation. 
However, in the case of dense gas ($N_e \ge 10^4$ cm$^{-3}$), collisional 
de-excitation of the H$_2$ molecule modifies the spectrum, approaching
the thermal spectrum. Excitation by hard X-ray are  ruled out 
by  other authors based on considerations of energetics \citep{RodriguezArdila04} 
and on the detection of the transition H$_2$ 2--1~S(3) in several Seyfert 
galaxies \citep{Davies05}, which would otherwise be suppressed in the case of 
 X-ray irradiated gas \citep{Davies05}.  
 
Another way to discriminate between the thermal and fluorescent 
excitations is through the rotational and vibrational temperatures.
In the case of thermal excitation both temperatures should be similar,
whereas in the case of fluorescent excitation a high vibrational temperature 
will be in contrast with a lower rotational temperature. We have 
determined both temperatures using the expressions given by \citet{Reunanen02}
using the H$_{2}$ line ratios 2--1~S(1)/1--0~S(1) and 
1--0S(0)/1--0S(2). The values obtained for this galaxy are $T_{\rm vib}$ $\la$ 4405 K and 
$T_{\rm rot}$ $\ga$ 1557 K, which are in good agreement 
with those derived by \cite{RodriguezArdila04} for a number of Seyfert 2
galaxies, such as Mrk~279 and NGC~5728. 
Hence, the most likely mechanism for the excitation of H$_2$ in Mrk~78 is 
UV fluorescent in a dense gas. Note, however, that, given the large uncertainties
in the line flux measurements, thermal excitation cannot be ruled out.

\subsection{Comparison with Photoionization Models}
\label{sc:cloudy}

We have used the photoinization code CLOUDY (version C05.07.06), 
in order to reproduce the line ratios obtained from our spectra 
(for a detailed description see \citet{Ferland03}). 
We computed a grid of models based on 
photoionization by a power-law continuum and physical conditions in the 
gas typical of NLR.
For the ionizing continuum we used a power-law shape of the form
$f_{\nu}\sim\nu^{\alpha}$, where $\alpha =-2$ is
the spectral index in the range above 10 \micron. 
At  energies  below  10 \micron\ the spectral index is 
$\alpha=5/2$. 
We assume a plane--parallel geometry, metallicity equal 0.3 solar, 
and grains with
properties similar to those of the Orion Nebula, which corresponds to the 
best-fitting models according to \citet{AlonsoHerrero}.
Three different values of hydrogen density were explored: $n_{\rm H}$ = 
10$^{4}$, 10$^{5}$, and 10$^{6}$~cm$^{-3}.$ 
An input to the code is the ionization parameter $U$, 
defined by $U= \mathrm{Q}_{\rm H} / (4\pi d^2 n_{\rm H} c)$.

\begin{figure}[!h]
\centering
\includegraphics[width=8cm]{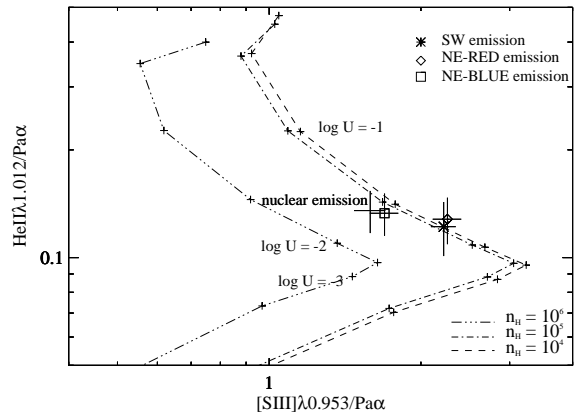}
\figcaption{\footnotesize{HeII$\lambda$1.012/Pa$\alpha$ versus 
[SIII]$\lambda$0.953/Pa$\alpha$ diagnostic diagram computed with the photoionization
code CLOUDY. Models are calculated for a power--law
continuum (index -2), different values of H density
(different line types) and a sequence of ionization
parameters. The marks along the U--sequence are 
separated by logU=0.5. Our measurements are 
also overplotted: cross for the
nucleus; asterisc for SW lobe; square for the 
blue component in the NE lobe; diamond for 
the red component in the NE lobe.} \label{ioniz1}}
\end{figure}

\begin{figure}[!h]
\centering
\includegraphics[width=8cm]{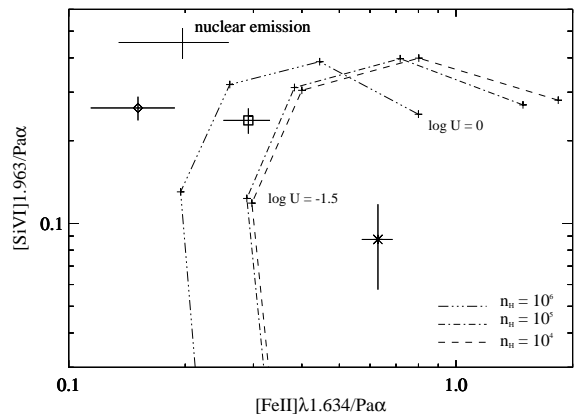}
\figcaption{\footnotesize{Same as Fig. \ref{ioniz1} but with 
[FeII]$\lambda$1.644/Pa$\alpha$ versus [SiVI]$\lambda$1.963/Pa$\alpha$ 
The meaning of symbols is the same as in Fig. \ref{ioniz1}}\label{ioniz2}}
\end{figure}

Line ratio diagrams are shown in Figs \ref{ioniz1} and \ref{ioniz2} 
for the following combination of lines: He II$\lambda$1.012/Pa$\alpha$ versus 
[S III]$\lambda$0.953/Pa$\alpha$ and [Si VI]$\lambda$1.963/Pa$\alpha$  versus 
 [Fe II]$\lambda$1.644/Pa$\alpha$.
Dashed lines represent the computed line ratios for the three hydrogen
densities employed, beginning with an ionization parameter of $\log U = 0 $
and decreasing it with a step of 0.5, towards the bottom of these plots.

Values of these line ratios from Table \ref{comps}, corresponding to 
the nucleus, the west and the east lobes (red and blue components for the latter), 
are overplotted for comparison with our simulations made with CLOUDY. 
It can be seen that data from our spectra are compatible with photoionization 
models for all regions considered within 
values of log $U$ in the range $-1$ to $-2$, except for the  
[Fe II]/Pa$\alpha$ line ratio.  
By looking at the diagram in Fig. \ref{ioniz1}, the best fit for
the nucleus and the east lobe--blue component is obtained when 
$\log{U} \sim -1.5.$
However, for the west lobe and east lobe--red component, 
the best fit is obtained when  
$\log{U} \sim -1.8.$ In all cases the H density is close to 
${n}_{\rm H} = 10^5\ {\mathrm cm}^{-3},$ which approaches the critical density 
for the [S III] line. 
From Fig. \ref{ioniz2}, it is easy to see that 
the  [Si VI]/Pa$\alpha$ line ratio is very sensitive to the value of the 
ionization parameter, becoming saturated at values $\log{U} \sim -0.5$.  
At higher values of the ionization parameter, 
the [Si VII]$\lambda 2.480$ transition will become more important. 
The nuclear emission can thus be explained with values $\log{U} \sim -0.5,$ 
whereas in the east lobe this value decreases to $\log{U} \sim -1.3,$ and 
in the west lobe to $\log{U} \sim -1.6.$ 
The differences between the best-fitting values of $U$ from the different line
ratios may be reconciled by lowering the spectral index of the power law, i.e., 
hardening the ionizing continuum, or using combination of clouds with different 
optical depths. 
In order to explain the [Fe II]/Pa$\alpha$ ratio, models may have to include 
effects due to interaction with the radio emission, such as increasing the iron abundance
after grain sputtering. 
   
We therefore conclude that the measured line ratios from the spectra of Mrk~78 
are compatible with photoinization by an AGN-like continuum.

\section{Conclusions}

We have presented and analyzed the NIR line spectrum of the 
nucleus and extended NLR of the Seyfert 2 galaxy Mrk~78.
The following results were found:  
\begin{itemize}
\item The nuclear spectrum and the extended NLR spectrum are 
produced by photoionization by a hard UV--X-ray continuum, as indicated by
the presence of the [Si VI] and [P II] lines. This fact is in good agreement
with the presence of a hidden active nucleus in Mrk 78. 
\item The nuclear emission in the $H$ band has numerous stellar absorption 
features.
According to the relative equivalent widths, the stellar population is 
dominated by K--M giants in this wavelength range. The nuclear stellar emission 
is diluted by a non-stellar component by a factor $\sim 34$\% in the $H$ band 
and $\sim 45$\% in the $K$ band.  
\item An enhancement of the [Fe II] emission is observed outside the
nucleus, in particular at $\sim 2\farcs$ to the west. This effect is interpreted
in terms of the interaction with the radio emission.  
\item Several transitions of the H$_2$ molecule are detected in the 
$K$ band. The dominant excitation mechanism could not be determined unambiguously,
although we favor UV fluorescence in a dense gas.  
\end{itemize}

\acknowledgments

We thank the anonymous referee for his very useful comments that led to
the improvement or our work.

C.R.A, J.A.P, R.B., \& A.M. acknowledge the Plan Nacional de Astronom\'{i}a y
      Astrof\'{i}sica (AYA2004-03136), which supported part of this work.

The William Herschel Telescope (WHT) is operated on the island of La Palma by 
the Isaac Newton Group in the Spanish Observatorio del Roque de los Muchachos 
of the Instituto de Astrof\'{i}sica de Canarias.

The image presented in Fig. 1 is based on observations made with the NASA/ESA 
{\it Hubble Space Telescope},
       obtained from the data archive at the Space Telescope Science Institute. 
       STScI is operated by the Association of Universities for Research in 
       Astronomy, Inc., under NASA contract NAS5-26555.

The authors acknowledge the data analysis facilities provided by the 
Starlink Project which is run by CCLRC on behalf of PPARC.


\begin{thebibliography}{}


\bibitem[Acosta-Pulido(2000)]{Acosta00} Acosta Pulido, J. A. 2000, 
ASPC, 216, 663

\bibitem[Acosta-Pulido et al.(2003)]{Acosta03} Acosta Pulido, J. A.,
et al. 2003, INGN, 7, 15

\bibitem[Alonso-Herrero et al.(1997)]{AlonsoHerrero} Alonso-Herrero, A., Rieke,
M. J., Rieke, G. H., \& Ruiz, M. 1997, \apj, 482, 747

\bibitem[Bautista \& Pradhan(1998)]{Bautista98} Bautista, M. A., \& Pradhan, A. K.
1998, \apj, 492, 650

\bibitem[Bertelli et al.(1994)]{Bertelli94} Bertelli, G., Bressan, A., Chiosi, C.,
Fagotto, F., \& Nasi, E. 1994, \aap, 106,275

\bibitem[Blietz et al.(1994)]{Blietz94} Blietz, M., et al. 1994, \apj, 421, 92

\bibitem[Capetti et al.(1994)]{Capetti94} Capetti, A., Macchetto, F., Sparks, 
W. B., \& Boksenberg, A. 1994, \apj, 421, 87

\bibitem[Cardelli et al.(1989)]{Cardelli89} Cardelli, J. A., Clayton, G. C., 
\& Mathis, J. S. 1989, \apj, 345, 245


\bibitem[Cecil et al.(2002)]{Cecil02} Cecil, G., Dopita, M. A., Groves, B., 
Wilson, A. S., Ferruit, P., P\'{e}contal, E., \& 
Binette, L. 2002, \apj, 568, 627

\bibitem[Cecil et al.(2000)]{Cecil00} Cecil, G., et al. 2000, \apj, 536, 675

\bibitem[Clements (1981)]{Clements81} Clements, E. 1981, \mnras, 197,829

\bibitem[Cutri et al.(2003)]{Cutri} Cutri, R. M., et al. 2003, 
VizieR On-line Data Catalog: II/246. Originally published in: 
University of Massachusetts and Infrared 
Processing and Analysis Center (IPAC/California Institute of Technology)
  
\bibitem[Dallier et al.(1996)]{Dallier96} Dallier, R., Boisson, C., \& Joly, M. 
1996, \aap, 116, 239

\bibitem[Davies et al.(2005)]{Davies05} Davies, R. I., Sternberg, A., Lehnert,
M. D., \& Tacconi-Garman, L. E. 2005, \apj, 633, 105

\bibitem[Doyon et al.(1994)]{Doyon94} Doyon, R., Joseph, R. D., \& Wright, G. S. 
1994, \apj, 421, 101 

\bibitem[Draine(1989)]{Draine89} Draine, B. T. 1989, isa, book, 93D

\bibitem[Draine \& Woods(1990)]{Draine90} Draine, B. T., \& Woods, D. T. 
1990, \apj, 363, 464

\bibitem[Ferland(2003)]{Ferland03} Ferland, G. J. 2003, \araa, 41, 517

\bibitem[Ferruit et al.(1999)]{Ferruit99} Ferruit, P., Wilson, A. S., Falcke, H., 
Simpson, C.,  P\'{e}contal, E., \& Durret, F. 1999, \mnras, 309, 1
 
 \bibitem[Ferruit(2002)]{Ferruit02} Ferruit, P. 2002, 
Emission Line from Jet Flows, 
ed. W. J. Henney, W. Steffen, A. C. Raga, \& L. Binette (Rev. Mexicana Astron.
Astrofis. Ser. Conf. 13)(M\'{e}xico, D.F.: Inst. Astron. Univ. Nac. Aut\'{o}noma 
M\'{e}xico), 183

  
\bibitem[F\"{o}rster Schreiber(2000)]{Forster00} F\"{o}rster Schreiber, N. M. 
2000, \aj, 120, 2089
  
\bibitem[Goodrich et al.(1994)]{Goodrich94} Goodrich, R. W., Veilleux, S., 
\& Hill, G. J. 1994, \apj, 422, 521

\bibitem[Hummer \& Storey(1987)]{hummer} Hummer, D. G., \& Storey, P. J. 
1987, \mnras, 224, 801
  
\bibitem[Ivanov et al.(2004)]{Ivanov04} Ivanov, V. D., Rieke, M. J., Engelbracht, 
C. W., Alonso-Herrero, A., Rieke, G. H., \& Luhman, K. L. 2004, \apjs, 151, 387
 
\bibitem[Ivanov et al.(2000)]{Ivanov00} Ivanov, V. D., Rieke, G. H., Groppi, E.,
Alonso-Herrero, A., Rieke, M. J., \& Engelbracht, C. W. 2000, \apj, 545, 190
    

\bibitem[Larkin et al.(1998)]{Larkin98} Larkin, J. E., Armus, L., Knop, R. A.,
Soifer, B. T., \& Matthews, K. 1998, \apjs, 114, 59
    
\bibitem[Levenson et al.(2001)]{Levenson01} Levenson, N. A., Weaver, K. A., 
\& Heckman, T. M. 2001, \apjs, 133, 269
    
\bibitem[Manchado et al.(2004)]{Manchado04} Manchado, A., et al. 2004, 
    in Proc. of the SPIE, 5492, 1094M
           
\bibitem[Marconi et al.(1994)]{Marconi94} Marconi, A., Moorwood, A.F.M.,
Salvati, M., \& Oliva, E. 1994, \aap, 291, 18
 
\bibitem[Michel \& Huchra(1988)]{Michel88} Michel, A., \& Huchra, J.
1988, \pasp, 100, 1423 
 
\bibitem[Mouri(1994)]{mouri} Mouri, H. 1994, \apj, 427, 777
 
 
\bibitem[Nelson \& Whittle(1996)]{Nelson96} Nelson, C. H., \& Whittle, M. 1996, 
 \apj, 465, 96

 
\bibitem[Neugebauer et al.(1979)]{neugebauer79} Neugebauer, G., Oke, J. B., 
Becklin, E. E., \& Mathews, K. 1979, \apj, 230, 79
        
         
\bibitem[Oliva \& Moorwood(1990)]{Oliva90} Oliva, E., \& Moorwood, A. F. M. 
1990, \apjl, 348, L5
 
\bibitem[Oliva et al.(1995)]{Oliva95} Oliva, E., Origlia, L., Kotilainen, J. K.,
\& Moorwood, A. F. M. 1995, \aap, 301, 55
 
\bibitem[Oliva et al.(2001)]{Oliva01} Oliva, E., et al. 2001, \aap, 369, L5
           
\bibitem[Origlia et al.(1993)]{Origlia93} Origlia, L., Moorwood A. F. M., \& 
Oliva, E. 1993, \aap, 280, 536
 
\bibitem[Origlia \& Oliva(2000)]{Origlia00} Origlia, L., \& Oliva, E. 2000,
      \aap, 357, 61
                     
\bibitem[Pedlar et al.(1989)]{pedlar} Pedlar, A., et al. 1989, \mnras, 238, 863

\bibitem[Pier \& Krolik(1992)]{Pier92} Pier, E. A., \& Krolik, J. H. 1992, \apj, 401,99

\bibitem[Prieto \& Viegas(2000)]{Prieto00} Prieto, M. A., \& Viegas, S. M.
2000, \apj, 532, 238

\bibitem[Quillen et al.(1999)]{Quillen99} Quillen, A. C., Alonso-Herrero, A.,
Rieke, M. J., Rieke, G. H., Ruiz, M., \& Kulkarni, V. 1999, \apj, 527, 696

\bibitem[Renzini \& Buzzoni(1986)]{Renzini86} Renzini, A., \&  Buzzoni, A. 
1986, in "Spectral Evolution of Galaxies", Chiosi C., Renzini A. (eds), p 195. 

\bibitem[Reunanen et al.(2002)]{Reunanen02} Reunanen, J., Kotilainen, J. K., 
  \& Prieto, M. A. 2002, \mnras, 331, 154

\bibitem[Reunanen et al.(2003)]{Reunanen03} Reunanen, J.,Kotilainen, J. K.,
\& Prieto, M. A. 2003, \mnras, 343, 192
      
\bibitem[Rhee \& Larkin(2005)]{Rhee05} Rhee, J. H., \& Larkin, J. E. 2005,
      \apj, 620, 151
    
\bibitem[Rodr\'\i guez-Ardila et al.(2004)]{RodriguezArdila04} Rodr\'\i guez-Ardila, A., Pastoriza, M. G.,
Viegas, S., Sigut, T. A. A., \& Pradhan, A. K. 2004, \aap, 425, 457

\bibitem[Rodr\'\i guez-Ardila et al.(2002)]{RodriguezArdila02} Rodr\'\i guez-Ardila, 
A., Viegas, S., Pastoriza, M. G., \& Prato, L.  2002, \apj, 519, 214

\bibitem[Schaller et al.(1992)]{Schaller92} Schaller, G., Schaerer, D., Meynet,
G., \& Maeder, A. 1992, \aap, 96, 269

\bibitem[Schlegel et al.(1998)]{Schlegel98} Schlegel, D. J., Finkbeiner, D. P., \&
Davis, M. 1998, \apj, 500, 525
      
\bibitem[Thatte et al.(1997)]{Thatte97} Thatte, N., Quirrenbach, A., Genzel, R., 
Maiolino, R., \& Tecza, M. 1997, \apj, 490, 238

\bibitem[Tran(1995)]{Tran95} Tran, H. D. 1995, \apj 440, 565
    
\bibitem[Vacca et al.(2003)]{vacca} Vacca, W. D., Cushing, M. C., \& Rayner, J. T.
 2003, \pasp, 115, 389
       
\bibitem[Veilleux  et al.(1997)]{Veilleux97} Veilleux, S., Goodrich R. W., 
Hill, G. J. 1997, \apj, 477, 631

\bibitem[Veilleux et al. (2002)]{Veilleux02} Veilleux, S., Cecil, G., 
Bland--Hawthorn, J., \& Shopbell, P.L. 2002, Emission Line from Jet Flows, 
ed. W. J. Henney, W. Steffen, A. C. Raga, \& L. Binette (Rev. Mexicana Astron.
Astrofis. Ser. Conf. 13)(M\'{e}xico, D.F.: Inst. Astron. Univ. Nac. Aut\'{o}noma 
M\'{e}xico), 222


\bibitem[Wallace \& Hinkle(1997)]{Wallace97} Wallace, L., \& Hinkle, K.
1997, \apjs, 111, 445
       
\bibitem[Whittle(1992)]{Whittle92} Whittle, M. 1992, \apj, 387, 109

\bibitem[Whittle et al.(1988)]{Whittle88} Whittle, M., Pedlar, A., Meurs,
E. J. A., Unger, S. W., Axon, D. J., \& Ward, M. J. 1988, \apj, 326, 125
       

\bibitem[Whittle \& Wilson(2004)]{Whittle04} Whittle, M., \& Wilson, A. S. 
2004, \aj, 127, 606
     
 
\end{thebibliography}
\end{document}